\documentclass[preprint2]{aastex6}
\usepackage{amsmath,amsfonts,amssymb}
\usepackage[version=3]{mhchem}
\usepackage[normalem]{ulem}
\newcommand{\myemail}{gnowak@iac.es}
\newcommand\chisq{\chi^{2}}

\newcommand\Mj{{\rm M_\mathrm{Jup}}}
\newcommand\Rj{{\rm R_\mathrm{Jup}}}

\newcommand\Ms{{\rm M_{\odot}}}
\newcommand\Rs{{\rm R_{\odot}}}

\newcommand\mps{{\rm m\,s^{-1}}}

\newcommand\kmps{{\rm km\,s^{-1}}}

\newcommand\gpcmcmcm{{\rm g\,cm^{-3}}}

\newcommand\maspyr{{\rm mas\,yr^{-1}}}
\newcommand\steff{T_\mathrm{eff}}
\newcommand\logsg{\log\,g_{\ast}}
\newcommand\sfeh{[Fe/H]}
\newcommand\sm{M_{\ast}}
\newcommand\sr{R_{\ast}}
\newcommand\sden{\rho_{\ast}}

\newcommand\svsini{{\it v}_\mathrm{rot} \sin\,i_{\star}}

\newcommand\svrot{{\it v}_\mathrm{rot}}
\newcommand\sprot{P_\mathrm{rot}}

\newcommand\pbm{M_\mathrm{b}}

\newcommand\pbr{R_\mathrm{b}}
\newcommand\pbden{\rho_\mathrm{b}}

\newcommand\vsini{$v$\,sin\,$i_\star$}

\newcommand\vmac{${\it v}_\mathrm{mac}$}

\newcommand\teff{$T_{\rm eff}$}

\newcommand\Msun{\hbox{$M_{\odot}$}}
\newcommand\Rsun{\hbox{$R_{\odot}$}}
\newcommand\kms{\hbox{km\,s$^{-1}$}}

\begin{document}

\title{EPIC\,219388192\,\lowercase{b} - an inhabitant of the brown dwarf desert \\ in the Ruprecht 147 open cluster}

\author{
Grzegorz Nowak\altaffilmark{1,2}, 
Enric Palle\altaffilmark{1,2}, 
Davide Gandolfi\altaffilmark{3,4}, 
Fei Dai\altaffilmark{5}, 
Antonino~F. Lanza\altaffilmark{6}, 
Teruyuki Hirano\altaffilmark{7}, 
Oscar Barrag\'{a}n\altaffilmark{3}, 
Akihiko Fukui\altaffilmark{8}, 
Hans Bruntt\altaffilmark{9}, 
Michael Endl\altaffilmark{10}, 
William~D. Cochran\altaffilmark{10}, 
Jorge Prieto-Arranz\altaffilmark{1,2}, 
Amanda Kiilerich\altaffilmark{9}, 
David Nespral\altaffilmark{1,2}, 
Artie~P. Hatzes\altaffilmark{11}, 
Simon Albrecht\altaffilmark{9},
Hans Deeg\altaffilmark{1,2}, 
Joshua~N. Winn\altaffilmark{12}, 
Liang Yu\altaffilmark{5}, 
Masayuki Kuzuhara\altaffilmark{13,14}
Sascha Grziwa\altaffilmark{15}, 
Alexis~M.~\,S. Smith\altaffilmark{16}
Pier~G. Prada Moroni\altaffilmark{17,18}, 
Eike~W. Guenther\altaffilmark{11}, 
Vincent Van Eylen\altaffilmark{19}, 
Szilard Csizmadia\altaffilmark{16}, 
Malcolm Fridlund\altaffilmark{20,21}, 
Juan Cabrera\altaffilmark{16}, 
Philipp Eigm\"uller\altaffilmark{16}, 
Anders Erikson\altaffilmark{16}, 
Judith Korth\altaffilmark{15}, 
Norio Narita\altaffilmark{22,13,14}, 
Martin P\"atzold\altaffilmark{15}, 
Heike Rauer\altaffilmark{16,23}, 
and Ignasi Ribas\altaffilmark{24}
}

\altaffiltext{1}{Instituto de Astrof\'\i sica de Canarias (IAC), 38205 La Laguna, Tenerife, Spain}
\altaffiltext{2}{Departamento de Astrof\'\i sica, Universidad de La Laguna (ULL), 38206 La Laguna, Tenerife, Spain}
\altaffiltext{3}{Dipartimento di Fisica, Universit\'a di Torino, Via P. Giuria 1, I-10125, Torino, Italy}
\altaffiltext{4}{Landessternwarte K\"onigstuhl, Zentrum f\"ur Astronomie der Universit\"at Heidelberg, K\"onigstuhl 12, D-69117 Heidelberg, Germany}
\altaffiltext{5}{Department of Physics and Kavli Institute for Astrophysics and Space Research, Massachusetts Institute of Technology, Cambridge, MA 02139, USA}
\altaffiltext{6}{INAF-Osservatorio Astrofisico di Catania, Via S.~Sofia, 78 - 95123 Catania, Italy}
\altaffiltext{7}{Department of Earth and Planetary Sciences, Tokyo Institute of Technology, 2-12-1 Ookayama, Meguro-ku, Tokyo 152-8551, Japan}
\altaffiltext{8}{Okayama Astrophysical Observatory, National Astronomical Observatory of Japan, Asakuchi, Okayama 719-0232, Japan}
\altaffiltext{9}{Stellar Astrophysics Centre, Department of Physics and Astronomy, Aarhus University, Ny Munkegade 120, DK-8000 Aarhus C, Denmark}
\altaffiltext{10}{Department of Astronomy and McDonald Observatory, Univerity of Texas at Austin, 2515 Speedway, Stop C1400, Austin, TX 78712, USA}
\altaffiltext{11}{Th\"{u}ringer Landessternwarte Tautenburg, Sternwarte 5, D-07778 Tautenburg, Germany}
\altaffiltext{12}{Princeton University, Department of Astrophysical Sciences, 4 Ivy Lane, Princeton, NJ 08544 USA}
\altaffiltext{13}{Astrobiology Center, National Institutes of Natural Sciences, 2-21-1 Osawa, Mitaka, Tokyo 181-8588, Japan}
\altaffiltext{14}{National Astronomical Observatory of Japan, 2-21-1 Osawa, Mitaka, Tokyo 181-8588, Japan}
\altaffiltext{15}{Rheinisches Institut f\"ur Umweltforschung an der Universit\"at zu K\"oln, Aachener Strasse 209, 50931 K\"oln, Germany}
\altaffiltext{16}{Institute of Planetary Research, German Aerospace Center, Rutherfordstrasse 2, 12489 Berlin, Germany}
\altaffiltext{17}{INFN, Section of Pisa, Largo Bruno Pontecorvo 3, I-56127, Pisa, Italy}
\altaffiltext{18}{Department of Physics "E. Fermi", University of Pisa, Largo Bruno Pontecorvo 3, I-56127, Pisa, Italy}
\altaffiltext{19}{Leiden Observatory, Leiden University, 2333CA Leiden, The Netherlands}
\altaffiltext{20}{Leiden Observatory, University of Leiden, PO Box 9513, 2300 RA, Leiden, The Netherlands}
\altaffiltext{21}{Department of Earth and Space Sciences, Chalmers University of Technology, Onsala Space Observatory, 439 92 Onsala, Sweden}
\altaffiltext{22}{Department of Astronomy, The University of Tokyo, 7-3-1 Hongo, Bunkyo-ku, Tokyo 113-0033, Japan}
\altaffiltext{23}{Center for Astronomy and Astrophysics, TU Berlin, Hardenbergstr. 36, 10623 Berlin, Germany}
\altaffiltext{24}{Institut de Ci\`encies de l'Espai (CSIC-IEEC), Carrer de Can Magrans, Campus UAB, 08193 Bellaterra, Spain}

\email{\myemail}

\shorttitle{EPIC\,219388192\,\lowercase{b}}
\shortauthors{Nowak et al.}
%

\begin{abstract}
We report the discovery of EPIC\,219388192\,b, a transiting brown dwarf in a 5.3-day orbit around a member star of Ruprecht-147, the oldest nearby open cluster association, which was photometrically monitored by K2 during its Campaign 7. We combine the K2 time-series data with ground-based adaptive optics imaging and high resolution spectroscopy to rule out false positive scenarios and determine the main parameters of the system. EPIC\,219388192\,b has a radius of $R_\mathrm{b}$=$0.937\pm0.042$~$\Rj$ and mass of $M_\mathrm{b}$=$36.50\pm0.09$~$\Mj$, yielding a mean density of $59.0\pm8.1$~$\gpcmcmcm$. The host star is nearly a Solar twin with mass $M_\star$=$0.99\pm0.05$~$\Ms$, radius $R_\star$=$1.01\pm0.04$~$\Rs$, effective temperature \teff=$5850\pm85$~K and iron abundance [Fe/H]=$0.03\pm0.08$~dex. Its age, spectroscopic distance, and reddening are consistent with those of Ruprecht-147, corroborating its cluster membership. EPIC\,219388192\,b is the first brown dwarf with precise determinations of mass, radius and age, and serves as benchmark for evolutionary models in the sub-stellar regime.
\end{abstract}


\section{Introduction}
Currently, more than one thousand brown dwarfs have been identified over the past 20 years, either isolated, in binary systems, or in orbit around more massive stars \citep[see][and references therein, as well as the DwarfArchives\footnote{\url{http://spider.ipac.caltech.edu/staff/davy/ARCHIVE/index.shtml}.}]{2016A&A...589A..49S}. In particular, the sample of brown dwarfs orbiting stars has increased in recent years thanks to exoplanet radial velocity (RV) surveys. The RV method enables the determination of the companion's orbital parameters and minimum mass $m\,\sin\,i$. With the assistance of the astrometric method, which allows the determination of the orbital inclination, the dynamical masses of several BDs have been measured \citep[e.g.][]{2011A&A...527A.140R,2016A&A...588A.144W}. Dynamical masses have also been measured for dozen or more brown dwarf binaries \cite[see, e.g., Table 1 in][and references therein]{2011hsa6.conf...48B}. However a model-independent and full characterization of the companion, i.e. the determination of its mass, radius and hence mean density is possible only for the eclipsing systems. 

The sample of eclipsing brown dwarfs with measured masses, radii, and densities known today contains 2 brown dwarf binaries -- namely 2MASS\,J05352184−0546085, an eclipsing binary system containing two extremely young brown dwarfs \citet{2006Natur.440..311S} and EPIC\,203868608\,b \citet{2016ApJ...816...21D} -- and 13 BDs that transit main sequence (MS) stars. The full list of eclipsing brown dwarfs, including the first 11 BDs transiting MS stars, is given in Table~1 of \cite{2016cole.book..143C}. The last two are the recently announced EPIC\,201702477\,b \citep{2016arXiv160604047B} and EPIC\,219388192\,b, the subject of this work. 

Here we present the discovery of the new eclipsing brown dwarf companion EPIC\,219388192\,b, which was observed by the Kepler K2 space mission during its Campaign 7. The uniqueness of EPIC\,219388192\,b stems from the fact that the host star is a member of the Ruprecht\,147 cluster \citep{2013AJ....145..134C}, providing a robust age determination. Based on the spectroscopic, as well as near-infrared and optical photometric isochrone fitting to the Dartmouth \citep{2008ApJS..178...89D}\footnote{\url{http://stellar.dartmouth.edu/models}.} and PARSEC \citep{2012MNRAS.427..127B}\footnote{\url{http://stev.oapd.inaf.it/cgi-bin/cmd}.} stellar evolution models, \cite{2013AJ....145..134C} determined an age of 2.75--3.25 Gyr for the Ruprecht\,147 cluster. Thus, a complete verification of the brown dwarf evolutionary models presented by \cite{2003A&A...402..701B} becomes possible for the first time. 

The paper is organized as follows: in Section~\ref{sec-observations} we describe the K2 data analysis and the complementary observations from the ground. In Section~\ref{sec-host_star} we describe the physical properties of the host star. In Section~\ref{sec-global_analysis} we describe the joint analysis of the radial velocity and photometric data. In Section~\ref{sec-tidal_evolution} we describe the tidal evolution of the system and in Section~\ref{sec-disasum} we provide a discussion and summary of our results.

\section{Observations and data reductions}
\label{sec-observations}

\subsection{K2 Photometry}

\begin{figure}[t]
\centerline{\includegraphics[angle=0,width=\linewidth]{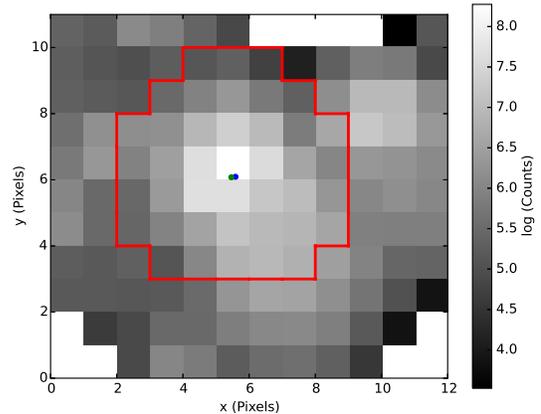}}
\caption{K2 image of EPIC\,219388192 with a customized aperture shown in red and defined based on the amount of light of each pixel and level of background light. The intensity of shading indicate the electron count, going from high (light grey) to low (dark grey).
\label{figure-01}}
\end{figure}

\begin{figure}[t]
\centerline{\includegraphics[angle=0,scale=0.125]{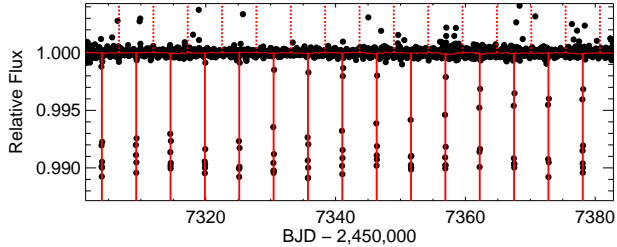}}
\caption{Detrended K2 light curve of EPIC\,219388192. The equally spaced vertical solid red lines mark the position of each transit.
\label{figure-02}}
\end{figure}

EPIC\,219388192 was a pre-selected target star of K2 Campaign 7 and, together with other 13\,550 target stars, was observed from the 4$^{th}$ of October to the 26$^{th}$ of December 2015. Images of EPIC\,219388192 were downloaded from the MAST archive\footnote{\url{https://archive.stsci.edu/k2/data\_search/search.php}.} and used to produce a detrended K2 light curve as described in detail in \cite{2016arXiv160901314D}. The pixel mask used to perform simple aperture photometry is presented in Figure~\ref{figure-01}. After extracting the time series data of all Field 7 targets, we searched the light curves for transiting planet candidates using the box fitting least-square (BLS) routine \citep{2002A&A...391..369K,2010ApJ...713L..87J} improved by implementing the optimal frequency sampling described in \cite{2014A&A...561A.138O}. The $\sim$1\,\%-deep transits of EPIC\,219388192\,b were clearly detected with a signal-to-noise ratio (SNR) of 15.8. A linear ephemeris analysis gave a best-fit period of 5.292569$\pm$0.000026 days and mid-time of the transit $T_{c,0}$=2457346.32942$\pm$0.00011 ($\mathrm{BJD_{TDB}}$). Figure~\ref{figure-02} shows the detrended light curve of EPIC\,219388192 with correction for centroid motions and baseline flux variations. The transit signals are marked with red lines. Table~\ref{table-03} reports the main identifiers of EPIC\,219388192 along with its coordinates, optical and near-infrared magnitudes and proper motion.

\subsection{High Contrast Imaging}
\label{subsec-hri-subaru_ircs+ao188}

We acquired high resolution, high contrast images of EPIC\,219388192 to search for potential nearby stars and estimate the contamination factor arising from these sources. We performed adaptive optics (AO) observations of EPIC\,219388192 on 19 June 2016 (UT) using the Subaru 188-elements Adaptive Optics system \citep[AO188;][]{2010SPIE.7736E..0NH} along with the Infrared Camera and Spectrograph \citep[IRCS;][]{2000SPIE.4008.1056K}. To search for faint nearby companions, we obtained $H$-band saturated images of EPIC\,219388192 with 5-point dithering and sidereal trucking. The exposure time was set to 15~sec. The sequence was repeated three times to increase the SNR. For each dithering position, we also obtained unsaturated frames of EPIC\,219388192 with individual exposure of 1.5 sec for the flux calibration.

The 15-sec exposure frames taken at 4 out of 5 dithering points reveal the presence of two faint objects South of EPIC\,219388192. To recover these faint stars, we discarded the frames in which these fainter stars were out of the field-of-view (FOV). Therefore, the total exposure time for the saturated images used for the subsequent analysis is 180~sec. On the other hand, these fainter stars were not visible in the 1.5-sec exposure frames, and hence we simply combined all the 5 unsaturated frames to measure the brightness of EPIC\,219388192.

Each image was dark-subtracted and flat-fielded in a standard manner. After the image distortion on each frame was corrected, the 12 saturated and 5 unsaturated images were respectively aligned and median-combined to create the final combined images. The FWHM of the stellar point-spread function (PSF) on the saturated and unsaturated images are $0\farcs1$ and $0\farcs09$, respectively.

\begin{table}[t]
\caption{Properties of companion candidates
\label{table-01}}
\tablewidth{0pt}
\begin{center}
\begin{tabular}{lrr}
\hline
\hline
\noalign{\smallskip}
Parameter   & SE Object   & SW Object \\
\noalign{\smallskip}
\hline
\noalign{\smallskip}
Separation ($^{\prime\prime}$) & $  5.998\pm0.012$ & $  7.538\pm0.015$ \\
Position Angle (deg)           & $142.740\pm0.060$ & $223.020\pm0.050$ \\
$\Delta m_H$ (mag)             & $  7.087\pm0.032$ & $  7.663\pm0.057$ \\
\noalign{\smallskip}
\hline
\end{tabular}
\end{center}
\end{table}

\begin{figure}[t]
\centerline{\includegraphics[angle=0,scale=0.475]{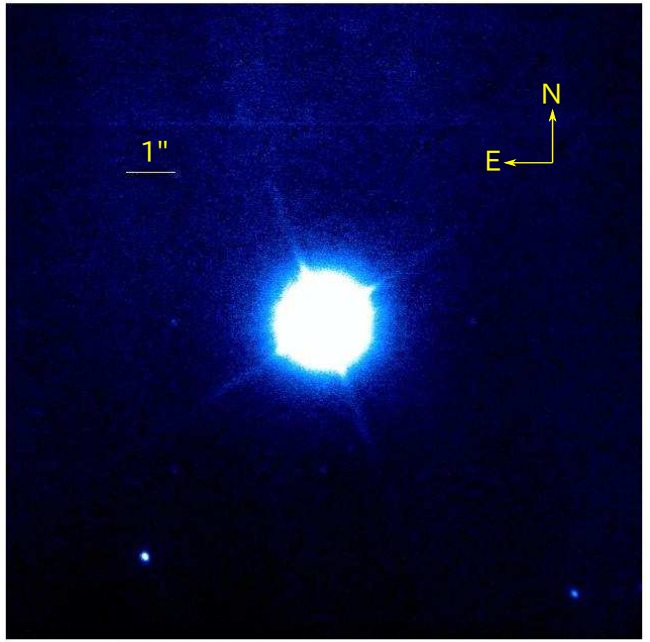}}
\caption{Combined saturated image of EPIC\,219388192 obtained with the Subaru/IRCS+AO188 instrument with FOV of $13^{\prime\prime}\times13^{\prime\prime}$.
\label{figure-03}}
\end{figure}

\begin{figure}[t]
\centerline{\includegraphics[angle=0,scale=1.150]{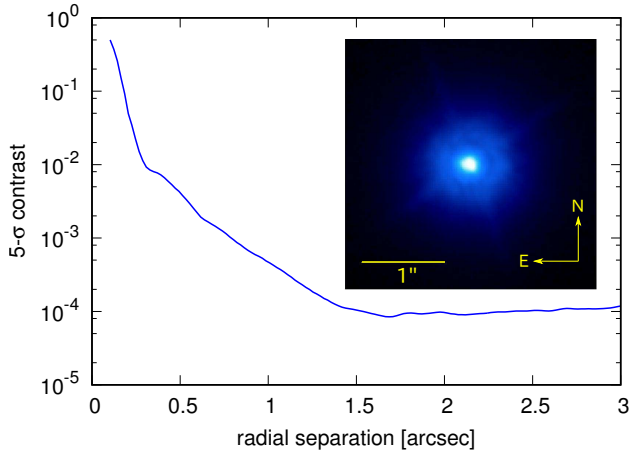}}
\caption{Five-$\sigma$ contrast curve as a function of angular separation from EPIC\,219388192. The inset displays the combined saturated image of the target with FOV of $3^{\prime\prime}\times 3^{\prime\prime}$.
\label{figure-04}}
\end{figure}

Figure \ref{figure-03} shows the combined, saturated image of EPIC\,219388192 with FOV of $13^{\prime\prime}\times13^{\prime\prime}$; the two faint stars are visible southwest (SW) and southeast (SE) of EPIC\,219388192. Table \ref{table-01} reports the separations, position angles, and $\Delta m_{H}$ of these 2 objects. The flux contrasts of these stars to EPIC\,219388192 ($<1.5\times 10^{-3}$) are much smaller than the observed K2 transit depth ($\sim 1\,\%$), implying that those cannot be sources of false positive signals. We also checked the inner region ($<1^{\prime\prime}$) around EPIC\,219388192 by visual inspection, but found no bright close-in companion (see the inset of Figure \ref{figure-04}). Following \citet{2016ApJ...820...41H}, we drew the 5-$\sigma$ contrast curve as a function of the angular separation from EPIC\,219388192, as shown in Figure \ref{figure-04}. 

\subsection{High Dispersion Spectroscopy}

\subsubsection{NOT/FIES}
We started the radial velocity (RV) follow-up of EPIC\,219388192 using the FIbre-fed {\'E}chelle Spectrograph \citep[FIES;][]{1999anot.conf...71F,2014AN....335...41T} mounted at the 2.56-m Nordic Optical Telescope (NOT) of Roque de los Muchachos Observatory (La Palma, Spain). We took 9 spectra between May and July 2016 as part of NOT observing programs P53-203, 53-109 and P53-016. We used the FIES high-resolution mode, which provides a resolving power of $R=67,000$ in the spectral range 3700--7300~{\AA}. Following the observing strategy described in \cite{2010ApJ...720.1118B} and \cite{2015A&A...576A..11G}, we traced the RV drift of the instrument by acquiring long-exposed ThAr spectra ($T_\mathrm{exp}$=35\,sec) immediately before and after each science exposure. The exposure time was set to 900-3600 sec according to weather conditions and observing schedule constraints. The data reduction follows standard IRAF and IDL routines, which include bias subtraction, flat fielding, order tracing and extraction, and wavelength calibration. Radial velocity measurements were computed via multi-order cross-correlations (CCF) with the RV standard star \object{HD\,50692} \citep{1999ASPC..185..367U} observed with the same instrument set-up as EPIC\,219388192. The SNR per pixel at 5500~\AA\ of the extracted spectra is in the range 15--35. Table~\ref{table-02} reports the FIES RVs, along with their 1-$\sigma$ error bars, CCF bisector spans (BS) and full-width half maximum (FWHM). Time stamps are given in Barycentric Julian Date in the Barycentric Dynamical Time \citep[BJD$_\mathrm{TDB}$; see, e.g.,][]{2010PASP..122..935E}.

\subsubsection{HJS/Tull}
We also observed EPIC\,219388192 with the Harlan J. Smith 2.7-m Telescope (HJS) and the Tull Coude Spectrograph \citep{1995PASP..107..251T} at McDonald Observatory (Texas, USA). The Tull spectrograph covers the entire optical spectrum at a resolving power of $R=60,000$. We obtained one spectrum of the star in June 2016 and two spectra in August 2016. We used exposures times of 1800~sec, which resulted in a SNR between 35 and 49 per resolution element at 5650~{\AA}. We calculated the absolute RV by cross-correlating the data with spectra of the RV-standard star \object{HD\,182488} (which we also observed in the same nights). Table~\ref{table-02} reports the extracted Tull RVs, along with their 1-$\sigma$ error bars.

\begin{table}
\caption{FIES and Tull RVs, CCF bisector spans and FWHMs.
\label{table-02}}
\begin{tabular}{lccrc}
\hline
\hline
BJD$_\mathrm{TDB}$ & RV & $\sigma_\mathrm{RV}$ & BIS      & FWHM \\
-$2,450,000$           & ($\mps$) & ($\mps$) &    ($\mps$) & ($\kmps$) \\     
\hline
\noalign{\smallskip}
\multicolumn{2}{l}{FIES} \\
7523.68062540   &   43713.500   &   32.663   &    15.5   &   12.999  \\
7525.61496665   &   49737.784   &   19.656   &    17.5   &   13.006  \\
7526.60509018   &   44979.980   &   18.852   &    -5.8   &   12.887  \\
7527.60734381   &   42396.504   &   21.930   &   -13.0   &   12.975  \\
7528.67908252   &   42872.233   &    9.904   &   -11.4   &   12.868  \\
7535.69323565   &   50637.688   &   15.878   &     4.7   &   13.035  \\
7566.63123022   &   46603.688   &   41.291   &    -8.5   &   12.796  \\
7567.60778355   &   50686.232   &   15.100   &   -14.8   &   12.936  \\
7568.52859679   &   46887.452   &   50.131   &   -67.5   &   12.949  \\
\hline
\noalign{\smallskip}
\multicolumn{2}{l}{Tull} \\
7543.80929600   &   41740.0     &   190.0    &     ---   &   ---  \\
7608.75108000   &   45210.0     &   200.0    &     ---   &   ---  \\
7609.70808000   &   49610.0     &   260.0    &     ---   &   ---  \\
\hline
\end{tabular}
\end{table}

\section{Properties of the host star}
\label{sec-host_star}

\subsection{Atmospheric and physical parameters}
\label{StellarParam}

We determined the photospheric parameters of EPIC\,219388192 from the co-added NOT/FIES spectra. The spectral analysis was performed with the versatile wavelength analysis {VWA} package\footnote{\url{https://sites.google.com/site/vikingpowersoftware/home}.} \citep{2012MNRAS.423..122B}. We measured an effective temperature \teff\,=\,$5850\pm85$~K, surface gravity $\logsg\,=\,4.38\pm0.12$~(cgs), and iron abundance \sfeh\,=\,$0.03\pm0.08$~dex. We adopted a macroturbulent velocity \vmac\,=\,3.4\,$\pm$0.6\,\kms\ \citep{Doyle2014} and measured a projected rotational velocity \vsini=4.1\,$\pm$\,0.4\,\kms\ by fitting the profile of many isolated and unblended metal lines.

The stellar mass, radius, and age were derived by combining \teff\ and \sfeh\ with the mean density $\rho_\star$ obtained from the transit light curve modeling (Section~\ref{sec-global_analysis}). We compared the position of  EPIC\,219388192 on a $\rho_\star$-versus-\teff\ with a grid of evolutionary tracks from the Pisa stellar evolution data base for low-mass stars\footnote{Available at \url{http://astro.df.unipi.it/stellar-models/}.} \citep{DellOmodarme2012}.

With a mass of $M_\star$=$0.99\pm0.05$~\Msun\ and radius of $R_\star$=$1.01\pm0.04$~\Rsun, EPIC\,219388192 is a Sun-like star. Stellar mass and radius imply a surface gravity of $\logsg=4.43\pm0.03$~(cgs), which agrees within 1-$\sigma$ with the value of $\logsg=4.38\pm0.12$~(cgs) derived from the NOT/FIES co-added spectra. We estimated an age of $3.9^{+1.9}_{-1.8}$~Gyr which is consistent with the Ruprecht 147 cluster's age of 2.75--3.25 Gyr determined by \citet{2013AJ....145..134C}.

We derived the interstellar extinction ($A_\mathrm{V}$) and distance ($d$) to the star following the technique outlined in \citet{Gandolfi2008}. Briefly, we fitted the magnitudes encompassed by the spectral energy distribution of the star to synthetic magnitudes extracted from the NEXTGEN model spectrum \citep{Hauschildt1999} with the same photospheric parameters as EPIC\,219388192. We adopted the extinction law of \citet{Cardelli1989} and assumed a normal total-to-selective extinction value of $R_\mathrm{v}$=3.1. We derived a reddening of $A_\mathrm{V}$\,=\,$0.35\pm0.05$~mag, which is consistent with the Ruprecht~147 cluster's extinction $A_\mathrm{V}$\,=\,$0.25\pm0.05$ measured by \citet{2013AJ....145..134C}. Assuming a black body emission at the star's effective temperature and radius, we measured a spectroscopic distance of EPIC\,219388192 $d$\,=\,$300\pm24$~pc, which is also in excellent agreement with the cluster's distance \citep[$d$\,=\,$295\pm15$~pc;][]{2013AJ....145..134C}.

\subsection{Stellar rotation and activity}

\begin{figure}[t]
\centerline{\includegraphics[angle=0,scale=0.125]{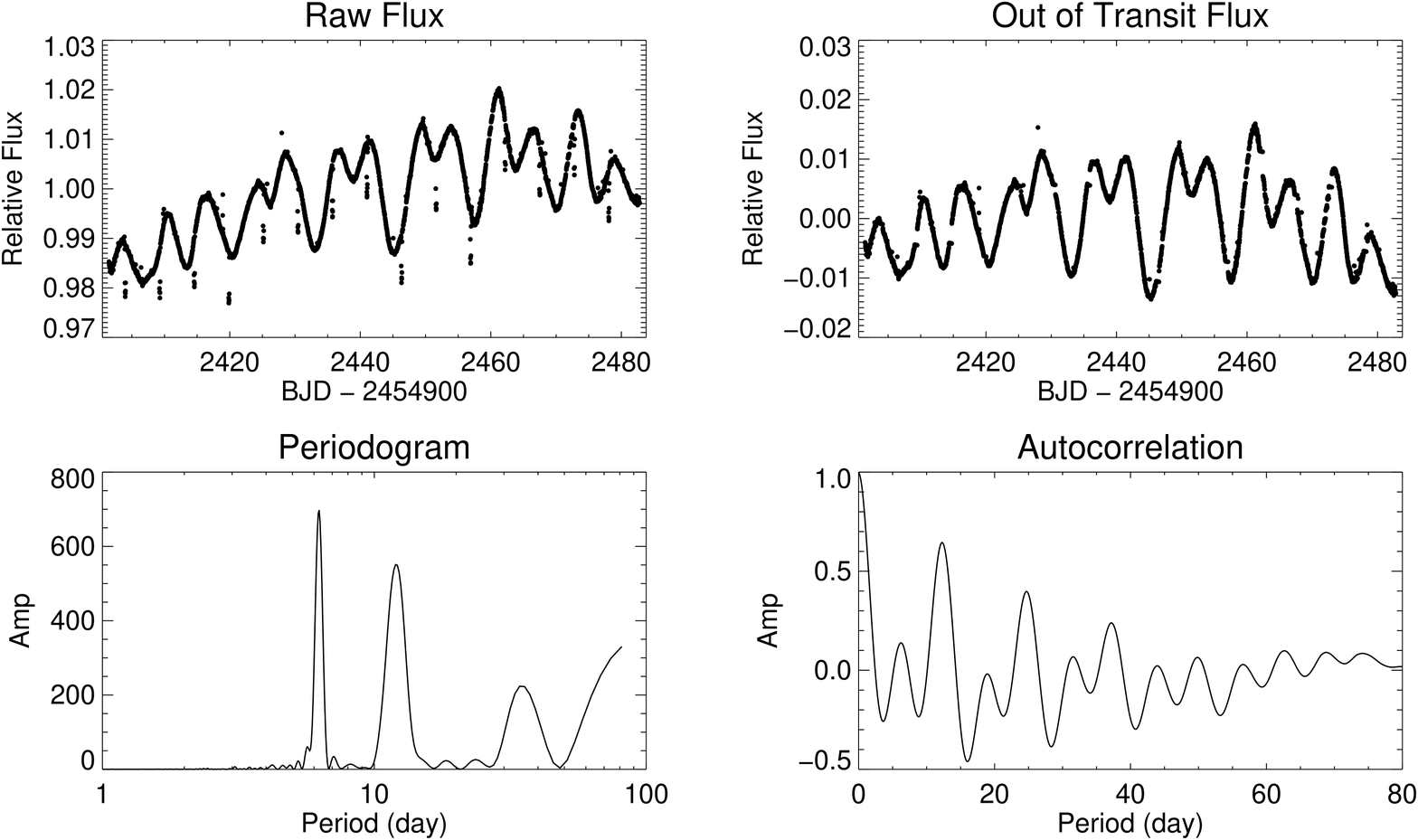}}
\caption{\emph{Upper left}: Raw flux of EPIC\,219388192 processed with a modified algorithm that better preserves stellar activity. \emph{Upper right}: Smoothly joined flux used for computing the Lomb-Scargle periodogram (bottom left panel) and auto cross-correlation function (bottom right panel).
\label{figure-05}}
\end{figure}

The light curve of EPIC\,219388192 displays periodic and quasi-periodic variations with a peak-to-peak photometric variation of about 2\%. Given the spectral type of the star, the observed variability is very likely ascribable to active regions (spots and faculae) carried around by stellar rotation. This is corroborated by the detection of emission components in the Ca H \& K lines. We measured the rotation period ($\sprot$) of EPIC\,219388192 using the auto-cross correlation function (ACCF) method \citep{McQuillan2014} applied to the out-of-transit light curve processed with a modified algorithm that better preserves stellar activity (Fig.\,\ref{figure-05}). The ACCF displays correlation peaks separated by $\sim$6.3~days, with a dominant peak at $\sim$12.6\,days (Fig.\,\ref{figure-05}). A visual inspection of the light curve reveals that features repeat every 12.6 days, suggesting that the latter is the rotation period of the star. The peaks occurring every 6.3~days are due to correlations between active regions at opposite stellar longitudes. We estimated a rotation period and uncertainty of $\sprot$\,=\,12.6\,$\pm$\,2.1 days defined as the position and the full width at half maximum of the strongest peak in the ACCF. The Lomb-Scargle periodogram shows also a significant peak at both the rotation period of the star and its first harmonic, corroborating our findings (Fig.\,\ref{figure-05}).

Our estimate of the projected rotational velocity (\vsini=4.1\,$\pm$\,0.4\,\kms; Sect.~\ref{StellarParam}) agrees with the equatorial velocity $\svrot = 2\pi\sr/\sprot$\,=\,4\,$\pm$\,1~$\kmps$ computed from the stellar radius $\sr$ and rotation period $\sprot$.

\begin{table}
\tabletypesize{\scriptsize}
\caption{Properties of EPIC\,219388192.
\label{table-03}}
\begin{center}
\begin{tabular}{lrr}
\hline
\hline
\noalign{\smallskip}
Parameter & Value & Source \\
\noalign{\smallskip}
\hline
\noalign{\smallskip}
\multicolumn{3}{c}{\emph{Coordinates and Main Identifies}}\\
\noalign{\smallskip}
RA 2000.0 (h)      & 19:17:34.036     & K2 EPIC\\
Dec 2000.0 (deg)   & -16:52:17.800    & K2 EPIC\\
2MASS Identifier   & 19173402-1652177 & 2MASS PSC\\
UCAC Identifier    & 366-166973       & UCAC4\\
\noalign{\smallskip}
\hline
\noalign{\smallskip}
\multicolumn{3}{c}{\emph{Optical and Near-Infrared Magnitudes}}\\
\noalign{\smallskip}
$Kepler$ (mag)  & 12.336                          & K2 EPIC\\
$B_{J}$ (mag)   & 13.284 $\pm$ 0.020              & K2 EPIC\\
$V_{J}$ (mag)   & 12.535 $\pm$ 0.020              & K2 EPIC\\
$g$ (mag)       & 12.854 $\pm$ 0.030              & K2 EPIC\\
$r$ (mag)       & 12.348 $\pm$ 0.020              & K2 EPIC\\
$i$ (mag)       & 12.348 $\pm$ 0.020              & K2 EPIC\\
$J$ (mag)       & 11.073 $\pm$ 0.023              & K2 EPIC\\
$H$ (mag)       & 10.734 $\pm$ 0.021              & K2 EPIC\\
$K$ (mag)       & 10.666 $\pm$ 0.021              & K2 EPIC\\
\noalign{\smallskip}
\hline
\noalign{\smallskip}
\multicolumn{3}{c}{\emph{Space Motion and Distance}}\\
\noalign{\smallskip}
PM$_\mathrm{RA}$ ($\maspyr$)           & -1.2 $\pm$ 1.4    & PPMXL\\
PM$_\mathrm{Dec}$ ($\maspyr$)          & -21.6 $\pm$ 3.4   & PPMXL\\
RV$_{\gamma,\mathrm{FIES}}$ ($\mps$)   & 45640 $\pm$ 10    & This work\\
RV$_{\gamma,\mathrm{Tull}}$ ($\mps$)   & 45840 $\pm$ 120   & This work\\
$d$ (pc) & 300 $\pm$ 24                & This work \\
$d$ (pc) & 295 $\pm$ 15                & 1\\
\noalign{\smallskip}
\hline
\noalign{\smallskip}
\multicolumn{3}{c}{\emph{Photospheric Parameters}}\\
\noalign{\smallskip}
$\steff$ (K)    & 5850 $\pm$ 85     & This work\\
$\logsg$ (dex)  & 4.38 $\pm$ 0.12   & This work\\
\sfeh\ (dex)    & 0.03 $\pm$ 0.08  & This work\\
\noalign{\smallskip}
\hline
\noalign{\smallskip}
\multicolumn{3}{c}{\emph{Derived Physical Parameters}}\\
\noalign{\smallskip}
$\sm$ ($\Ms$)           & 0.99 $\pm$ 0.05                & This work\\
$\sr$ ($\Rs$)           & 1.01 $\pm$ 0.04                & This work\\
Age (Gyr)               & $3.9^{+1.9}_{-1.8}$            & This work \\
Age (Gyr)               & 2.75 -- 3.25                   & 1\\
\noalign{\smallskip}
\hline
\noalign{\smallskip}
\multicolumn{3}{c}{\emph{Stellar Rotation}}\\
\noalign{\smallskip}
$\sprot$ (days)         & 12.6 $\pm$ 2.10               & This work\\
$\svsini$ ($\kmps$)     & 4.1 $\pm$ 0.4                  & This work\\
\noalign{\smallskip}
\hline
\end{tabular}
\end{center}
\vspace{-0.5cm}
\tablecomments{1: from \citet{2013AJ....145..134C}.}
\end{table}

\section{Global Analysis}
\label{sec-global_analysis}

To estimate the system parameters, we performed a global joint analysis of the K2 transit light curves and NOT/FIES and HJS/Tull radial velocity measurements using the following $\chisq$ statistic:
\begin{align}
\chisq = \sum_\mathrm{i=1}^{i=N_\mathrm{f}}\frac{\left(f_\mathrm{obs,i}-f_\mathrm{mod,i}\right)^{2}}{\sigma_\mathrm{f,i}^{2}}\notag\\
       + \sum_\mathrm{i=1}^{i=N_\mathrm{FIES,RV}}\frac{\left(RV_\mathrm{FIES,obs,i}-RV_\mathrm{FIES,mod,i}\right)^{2}}{\sigma_\mathrm{FIES,RV,i}^{2}}\\
       + \sum_\mathrm{i=1}^{i=N_\mathrm{Tull,RV}}\frac{\left(RV_\mathrm{Tull,obs,i}-RV_\mathrm{Tull,mod,i}\right)^{2}}{\sigma_\mathrm{Tull,RV,i}^{2}}\notag,
\end{align}
where $N_\mathrm{f}$, $N_\mathrm{FIES,RV}$, and $N_\mathrm{Tull,RV}$ are the number of the K2 photometric, NOT/FIES, and HJS/Tull radial velocity measurements respectively, and $f_\mathrm{obs,i}$, $RV_\mathrm{FIES,obs,i}$, and $RV_\mathrm{Tull,obs,i}$ are $i-$th observed K2 flux, NOT/FIES and HJS/Tull RV, and finally $\sigma_\mathrm{f,i}$, $\sigma_\mathrm{FIES,RV,i}$ and $\sigma_\mathrm{Tull,RV,i}$ are their errors. For the RV model we adopted the following equations:
\begin{equation}
RV_\mathrm{FIES,mod,i} = K\left[\cos\left(\nu+\omega\right)+e\cos\left(\omega\right)\right] + \gamma_\mathrm{FIES},
\end{equation}
\begin{equation}
RV_\mathrm{Tull,mod,i} = K\left[\cos\left(\nu+\omega\right)+e\cos\left(\omega\right)\right] + \gamma_\mathrm{Tull},
\end{equation}
where $K$ is the RV semi-amplitude, $\nu$ is the true anomaly, $\omega$ is the argument of periastron, $e$ is the eccentricity, $\gamma_\mathrm{FIES}$ is the systemic velocity as measured from the NOT/FIES RV measurements, and $\gamma_\mathrm{Tull}$ is the systemic velocity as measured from the HJS/Tull RV measurements. For the transit model, we used the Python package \texttt{BATMAN} \citep{2015PASP..127.1161K} to calculate the light curve. 

There are 4 global parameters in our joint fit: time of conjunction ($T_\mathrm{c}$), orbital period ($P_\mathrm{orb}$), eccentricity ($e$), and argument of pericenter ($\omega$). To avoid the bias towards non-zero eccentricity \citep{1971AJ.....76..544L}, we transformed $e$ and $\omega$ to $\sqrt{e} \cos\,\omega$ and $\sqrt{e} \sin\,\omega$ during the fitting. There are five additional parameters involved in producing the light curve: cosine of orbital inclination ($\cos\,i$), radius ratio ($\pbr/\sr$), semi-major axis in units of stellar radius ($a/\sr$), and the quadratic limb darkening coefficients ($u_{1}$ and $u_{2}$). In the Keplerian model we fit the stellar jitter ($\sigma_{j}$). Uniform priors were adopted for all parameters.

\begin{figure}[t]
\centerline{\includegraphics[angle=0,width=\linewidth]{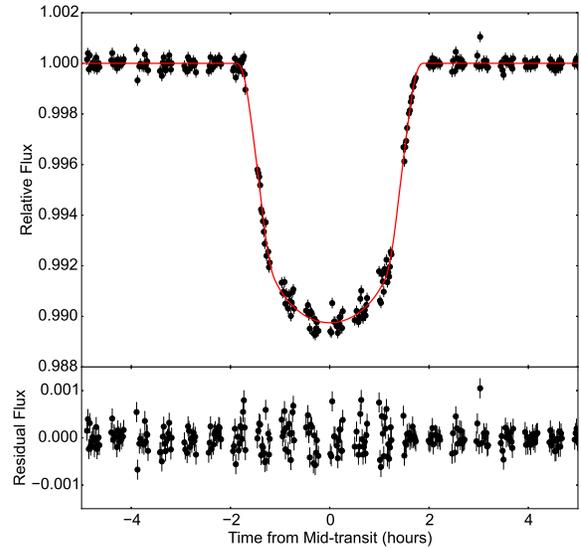}}
\caption{\emph{Upper panel}. EPIC\,219388192's transit light curves folded to the orbital period of the planet and best-fitting transit model (red line). \emph{Lower panel}. Residuals to the fit.
\label{figure-06}}
\end{figure}

\begin{figure}[t]
\centerline{\includegraphics[width=\linewidth]{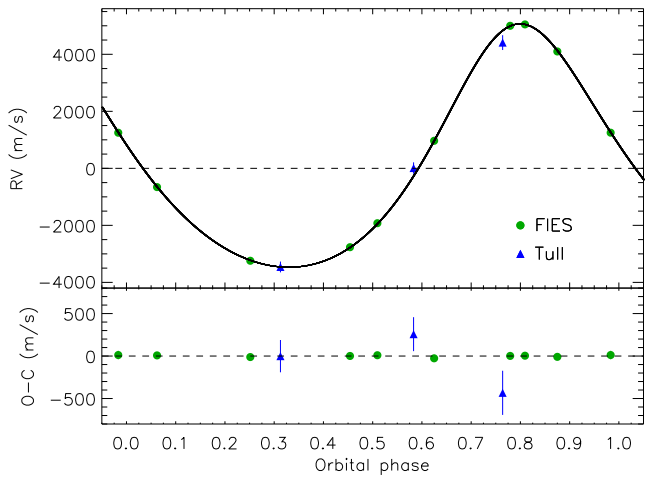}}
\caption{\emph{Upper panel}. Phase folded FIES (green circles) and Tull (blue triangles) RVs of EPIC\,219388192 and best fitting Keplerian model (thick line). \emph{Lower panel}. RV residuals to the fit.
\label{figure-07}}
\end{figure}

\begin{table*}
\tabletypesize{\scriptsize}
\caption{Results from the global fit of the photometric and spectroscopic data of EPIC\,219388192.
\label{table-04}}
\tablewidth{0pt}
\begin{center}
\begin{tabular}{lr}
\hline
\hline
\noalign{\smallskip}
Parameter                                                         & Value \\
\noalign{\smallskip}
\hline
\multicolumn{2}{c}{\emph{Fitted parameters}}\\
\noalign{\smallskip}
Orbital period $P_\mathrm{orb}$ (days)                            & 5.292569 $\pm$ 0.000026\\
Epoch of the transit $T_\mathrm{0,b}$ (BJD$\mathrm{_{TDB}}$)      & 2457346.32942$\pm$0.00011\\
Scaled radius $\pbr/\sr$                                          & 0.09321 $\pm$ 0.00046\\
Scaled semi-major axis $a/\sr$                                    & $12.62^{+0.10}_{-0.15}$\\
Orbit inclination $i$ (degrees)                                   & 90.0 $\pm$ 0.7\\
Impact parameter $b$                                              & 0.00 $\pm$ 0.15\\
Linear limb darkening coefficient $u_{1}$                         & 0.468 $\pm$ 0.040\\
Quadratic limb darkening coefficient $u_{2}$                      & 0.013 $\pm$ 0.087\\
Orbit Eccentricity $e$                                            & 0.1929 $\pm$ 0.0019\\
Stellar argument of periastron $\omega$                           & 345.9 $\pm$ 1.0\\
RV semi-amplitude variation $K$ ($\mps$)                          & 4267 $\pm$ 12\\
Systemic velocity $\gamma_{\mathrm{FIES}}$ ($\mps$)               & 45640 $\pm$ 10\\
Systemic velocity $\gamma_{\mathrm{Tull}}$ ($\mps$)               & 45840 $\pm$ 120\\
RV jitter $\sigma_{j}$ ($\mps$)                                   & $9^{+13}_{-6}$ \\
\noalign{\smallskip}
\hline
\noalign{\smallskip}
\multicolumn{2}{c}{\emph{Derived parameters}}\\
\noalign{\smallskip}
Brown dwarf mass $\pbm$ ($\Mj$)                                   & 36.50 $\pm$ 0.09\\
Brown dwarf radius $\pbr$ ($\Rj$)                                 & 0.937 $\pm$ 0.042\\
Brown dwarf mean density $\pbden$ ($\gpcmcmcm$)                   & 59.0 $\pm$ 8.1\\
Brown dwarf equilibrium temperature (K)$^{1}$                     & 1164 $\pm$ 40\\
Semi-major axis $a$ (au)                                          & 0.0593$\pm$0.0029\\ 
Host star mean density $\sden$ ($\gpcmcmcm$)                      & 1.369 $\pm$ 0.056\\
\noalign{\smallskip}
\hline
\end{tabular}
\end{center}
\vspace{-0.5cm}
\tablecomments{1: Assuming 	isotropic reradiation and a Bond albedo of zero.}
\end{table*}

We first obtained the best-fit solution using the Levenberg-Marquart algorithm as implemented in the \verb+lmfit+ package in python. To obtain the uncertainties and covariances on various parameters, we performed a MCMC analysis using the Python package \verb+emcee+ \citep{2013PASP..125..306F}. We started 250 walkers drawn from a Gaussian distribution in parameter space, centered on the minimum-$\chi^2$ solution. We stopped the walkers after 5000 links. We then checked the convergence by calculating the Gelman-Rubin potential scale reduction factor \citep{gelman_rubin-1992-statistical_science-7-457-511-inference_from_iterative_simulation_using_multiple_sequences} dropped below 1.02. We reported the median and the 16\% and 84\% percentiles of the marginalized posterior distribution for each parameters in Table~\ref{table-04}. The observed data along with the best-fit models are displayed in Figures \ref{figure-06}--\ref{figure-07}, for the phase-folded K2 light curve and orbital RVs, respectively. To check our results, we also modeled the data with the code \texttt{pyaneti} (Barrag\'an et al., in prep.). The parameter estimates are in agreement well within 1-$\sigma$.

The joint analysis allows the orbital configuration to be constrained to high precision. The orbit is relatively eccentric $e$=0.1929$\pm$0.0019. The joint analysis also derived a stellar density of 0.97$\pm$0.04 solar density. The residual fluxes within the transit window show a larger scatter than those out of the transit window. We interpret this as the result of spot crossing anomalies: when the planet occults a star spot during a transit, the planet occults a dimmer part of the stellar photosphere and therefore the observed flux will be higher than expected.

\section{Tidal evolution of the system}
\label{sec-tidal_evolution}
\begin{figure}[t]
\centering{\includegraphics[width=\linewidth]{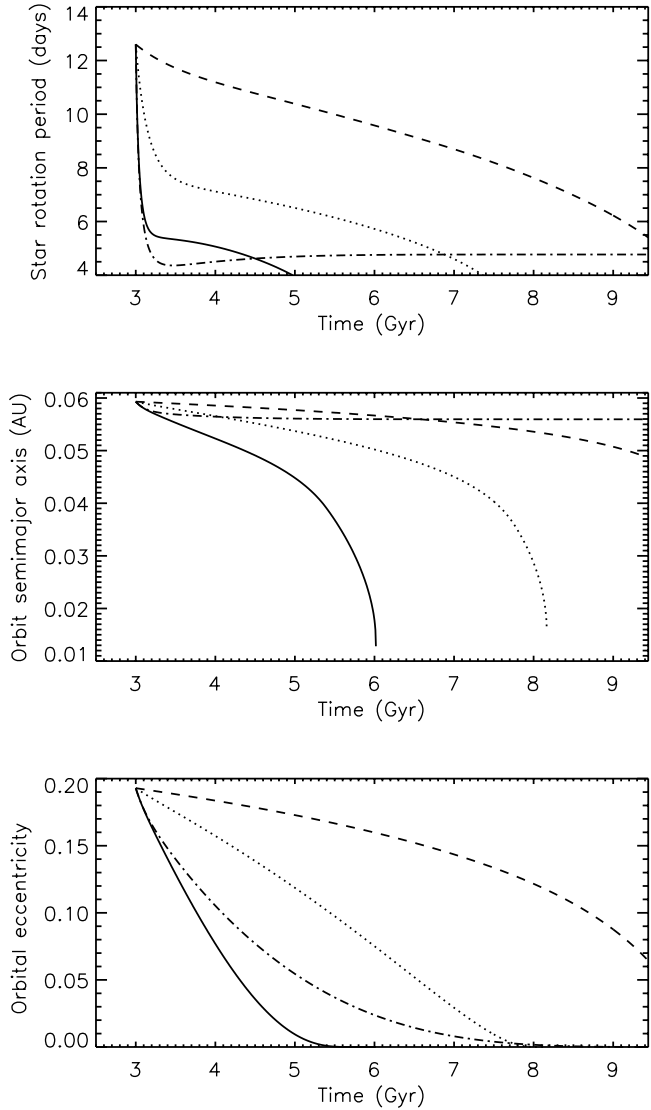}}
\caption{Upper panel: Evolution of the  stellar rotation period for $Q^{\prime}_{*} = 2.0 \times 10^{6}$ (solid line), $Q^{\prime}_{*} = 10^{7}$ (dotted line), and $Q^{\prime}_{*} = 5 \times 10^{7}$ (dashed line); the case without wind braking and $Q^{\prime}_{*} = 2 \times 10^{6}$ is also shown for comparison (dash-dotted line). Middle panel: as in  the upper panel, for the evolution of the orbital semi-major axis. Lower panel: as in the upper panel, for the evolution of the eccentricity.
\label{figure-08}}
\end{figure}

\begin{figure}[t]
\centering{\includegraphics[width=\linewidth]{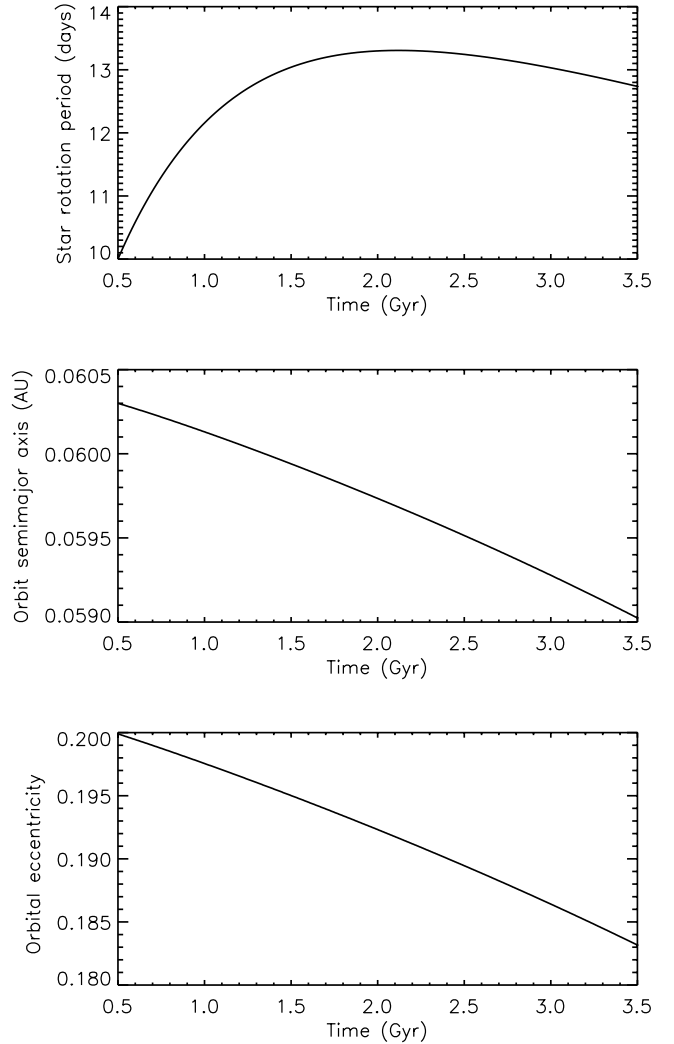}}
\caption{The same as Fig.~\ref{figure-08}, but assuming a rotation period of  10 days when the star had an age of 500~Myr and $Q^{\prime}_{*} = 7 \times 10^{7}$.
\label{figure-09}}
\end{figure}

EPIC\,219388192 is an interesting system to study tidal interactions between a brown dwarf and a main-sequence star. Assuming that the modified tidal quality factors of the star and the brown dwarf are similar (see below), most of the tidal kinetic energy is dissipated inside the star rather than inside the brown dwarf because the ratio $\eta \equiv (\rho_{\rm b}/\rho_{*})^{2} (R_{\rm b}/R_{*}) \simeq 157.5 \gg 1 $ \citep[cf.][]{Ogilvie14}. The ratio $\zeta$ of the present orbital angular momentum to the stellar spin angular momentum is $\zeta \sim 150$ assuming a gyration ratio of the star equal to that of the Sun at an age of 3~Gyr, i.e., $\beta = 0.289$ \citep{Claret04}. This implies that the tidal evolution of the stellar spin proceeds remarkably faster than that of the orbit with a transfer of angular momentum from the orbit to the stellar rotation because $P_{\rm rot} > P_{\rm orb}$ \citep{Ogilvie14}. Indeed, we find that the rotation period of the star $P_{\rm rot}$ is significantly shorter than expected on  the basis of gyrochronology because, by applying Eq.~(3)  of \citet{Barnes07}, we estimate a rotation period of $\sim 18.7$~days for a single  sun-like star of $\sim 3$~Gyr of age. 

A preliminary model of the tidal evolution of the system is computed according to the approach of  \citet{Leconteetal10} that we modify by considering constant modified tidal quality factors for the star and the brown dwarf indicated with $Q^{\prime}_{*}$ and $Q^{\prime}_{\rm b}$, respectively. They are related to the constant time lag of the tides inside the corresponding body by means of Eq.~(19) of \citet{Leconteetal10}. Note that a smaller value of $Q^{\prime}$ implies an higher  dissipation rate of the tidal energy inside the body. Moreover, we add the angular momentum loss produced by the stellar magnetized wind by considering a Skumanich-type law with saturation at an angular velocity equal to eight times that of the present Sun  \citep[e.g., Eq.~2 in][]{Spadaetal11} and assume a rigidly rotating star the radius of which changes in time according to a 1~M$_{\odot}$ model \citep{DellOmodarme2012}.

The evolution of the system parameters is plotted in Fig.~\ref{figure-08} for different values of $Q^{\prime}_{*}$ ranging from $2.0\times 10^{6}$ to $5 \times 10^{7}$; for comparison, we plot also the evolution for $Q^{\prime}_{*} = 2.0 \times 10^{6}$ without any wind braking. The orbital angular momentum and the stellar spin are assumed to be aligned with a present age of the system of 3~Gyr. The current ratio of the stellar rotation period to the orbital period is close but still above the critical value $P_{\rm rot}/P_{\rm orb} = 2$ for the excitation of inertial waves inside the star that would remarkably increase tidal dissipation \citep{OgilvieLin07}. Since the star is spun up by tides, the critical value for the excitation of those waves is predicted to be reached within the next few hundred Myr for $Q^{\prime} \leq 10^{7}$, while $\sim 2$~Gyr will be required for $Q^{\prime}_{*} = 5 \times 10^{7}$   due to the slower acceleration of the stellar rotation. Beyond that threshold, the value of $Q^{\prime}_{*}$ will remarkably decrease accelerating the tidal evolution. In our constant-$Q^{\prime}$ approximation, this would favour the model computed with the smallest value of $Q^{\prime}_{*}$ with a fast spin up of the star followed by the orbital decay of the system within $\approx$3~Gyr. The spin evolution is faster than the orbital decay because $\zeta \gg 1$. If the wind braking were absent, the system would avoid the orbital decay with the star reaching synchronization at a rotation period of $\sim$4.7~days and the orbit becoming circular with only a slight decrease of the semimajor axis thanks to the large reservoir of angular momentum in the present orbit. The decay of the system is therefore a consequence  of the magnetic wind braking with a phase of reduced acceleration of the stellar spin when the tidal spin up and the wind loss temporarily balance with each other \citep{DamianiLanza15}. The increase of the stellar radius along the main-sequence increases slightly the synchronization period, but does not affect our results. 

The past evolution of the system is much more uncertain because we have no idea of its initial conditions. We may assume that the tidal interaction was not strong in the past because the rotation period of the star was too long for the excitation of inertial waves. As an illustrative model, we plot in Fig.~\ref{figure-09} the evolution with $Q^{\prime} = 7 \times 10^{7}$ and a rotation period of 10~days at the age of 500 Myr when our  model assuming a rigid internal  rotation becomes to be applicable. This is the typical rotation period of  slowly rotating single stars of 1~M$_{\odot}$ in an open cluster of that age \citep[cf.][]{GalletBouvier15}. We see that the wind braking is initially stronger than the tidal spin up, but when the star reaches an age of $\sim 2$~Gyr, the tidal torque becomes dominant and the evolution of the stellar spin is reversed. The decay of the semi-major axis and of the  eccentricity is very small because $\zeta \gg 1$, suggesting that the present eccentricity could be a remnant of the formation phase of the system.

The above results are weakly dependent on the value of $Q^{\prime}_{\rm b}$, that we assume to be $10^{6}$ in all our calculations, because $\eta \gg 1$. The rotation of the brown dwarf is rapidly synchronized with the orbital motion within $0.1-10$~Myr for a wide range of $Q^{\prime}_{\rm b}$ \citep[cf.][]{Leconteetal10}, thus we assume it is rotating synchronously since the beginning in all our calculations.

\section{Discussion and Summary}
\label{sec-disasum}

\subsection{Ruprecht 147 cluster membership}
\label{subsec-r147_mem}

The EPIC\,219388192's membership probability to the Ruprecht 147 cluster was reported by \cite{2013AJ....145..134C} as ``possible''. This was motivated by the radial velocity of EPIC\,219388192 measured by the authors to be 47.3~$\kmps$. This value is $\sim$6~$\kmps$ higher than the cluster's average RV, $40.86\pm0.56$~$\kmps$, which was determined by \cite{2013AJ....145..134C} based on the RV measurements of six known cluster members. The systemic velocity of EPIC\,219388192 as measured using the NOT/FIES and HJS/Tull spectra is equal to 45.640$\pm$0.010~$\kmps$ and 45.840$\pm$0.120~$\kmps$, respectively, i.e., $\sim$2~$\kmps$ lower than the value measured by \cite{2013AJ....145..134C}. One possible reason of this discrepancy is the high $K$ semi-amplitude of EPIC\,219388192\,b. The other one may be the systematic shifts of the RV offsets between different spectrographs.

On the other hand, our estimates of the distance, reddening and age of EPIC\,219388192 (Sect.~\ref{StellarParam}) are all consistent with those of Ruprecht 147. We conclude there is now solid evidence for the star being a member of the Ruprecht 147 cluster.

In section~\ref{subsec-hri-subaru_ircs+ao188} we present the detection of two faint stars close to EPIC\,219388192. If we assume that the two objects are members of Ruprecht 147, we can obtain further information on these stars. Adopting the cluster's distance of $295\pm15$\,pc, the angular separations  imply a distance of $1769\pm90$ au (SE object) and $2224\pm113$ au (SW object) between EPIC\,219388192 and the two sources. The apparent magnitude $m_H=10.734\pm 0.021$~mag of EPIC\,219388192 yields an absolute magnitude of $M_H=3.38\pm 0.11$~mag. Thus, the magnitude differences listed in Table \ref{table-01} translate into absolute magnitudes of $M_H=10.47\pm 0.12$ mag (SE object) and $M_H=11.05\pm 0.13$ mag (SW object). According to the Dartmouth isochrone table \citep{2008ApJS..178...89D}, such faint stars ($M_H$\,$>$\,10 mag) would be very late-type M dwarfs (later than M8) or brown dwarfs, with their masses being less than $\sim$0.1~$M_\odot$. It would be of great interest if such multiple late-type stars, including EPIC\,219388192\,b, are clustered within a relatively small region. Further observations (e.g., adaptive optics imaging in different bands) are required to verify the memberships of those faint objects.

\begin{figure}[t]
\centerline{\includegraphics[angle=0,width=\linewidth]{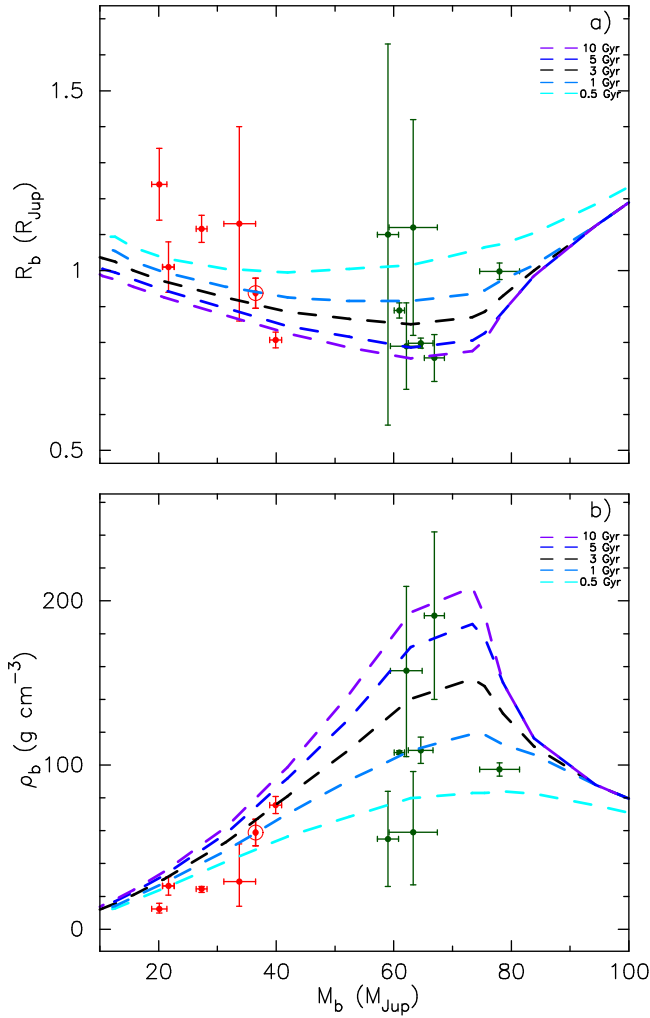}}
\caption{The mass -- radius (a)) and mass -- density (b)) relationships for all brown dwarfs transiting MS stars. The red filled circles indicate brown dwarfs with masses below 45~$\Mj$ at which \cite{2014MNRAS.439.2781M} report a gap in the mass distribution. The dark-green filled circles indicate brown dwarfs with masses above 45~$\Mj$. EPIC\,219388192\,b is indicated as a red filled circle with a rim. The dashed lines indicates the COND03 model radii and densities for brown dwarfs of 10 Gyr (violet), 5 Gyr (blue), 3 Gyr (black), 1 Gyr (light-blue) and 0.5 Gyr (cyan). Based solely on the fit to these models EPIC\,219388192\,b would seem to have an age of 1 Gyr. \label{figure-10}}
\end{figure}

\subsection{Comparison with the \cite{2003A&A...402..701B} COND03 models}

According to the COND03 evolutionary models for cool sub-stellar objects \citep{2003A&A...402..701B}, a 3-Gyr-old brown dwarf with a mass of 36.5~$\Mj$ should have a radius of 0.9015~$\Rj$ and a mean density of 69.71~$\gpcmcmcm$. Our estimates of the radius and density of EPIC\,219388192\,b are $R_\mathrm{b}$\,=\,$0.937\pm0.042$\,$\Rj$ and $\rho_\mathrm{b}$\,=\,$59.0\pm8.1$~$\gpcmcmcm$. Within 1-$\sigma$ they agree with the values expected from the COND03 models. However most of brown dwarfs known to transit MS stars seem to be inflated (Figure~\ref{figure-10}). Commonly proposed mechanisms to explain inflated exoplanets, like their host star irradiation, tides, increased interior opacity or efficiency of the heat transfer, have little effect on brown dwarfs that are considerably more massive than exoplanets \citep[see, e.g.,][]{2011A&A...533A..83B,2011A&A...525A..68B}. The increase in the atmospheric opacity proposed by \cite{2007ApJ...661..502B} that accounts for the slower cooling is currently thought to be mainly responsible for radius anomalies of transiting BDs. To verify this hypothesis, measurements of a secondary eclipse of EPIC\,219388192\,b in the infrared would be highly desirable. Thanks to the known effective temperature from COND03 models (800~K) and equilibrium temperature ($T_\mathrm{eq}$\,=\,1164$\pm$40~K), EPIC\,219388192\,b is also a benchmark for testing the effects of stellar irradiation. The eccentric, short-period orbit with well-known age makes also EPIC\,219388192\,b an excellent - and unique - candidate to check the theories of star -- brown dwarf tidal interactions in the presence of magnetic stellar winds \citep[cf.][]{2015ApJ...807...78F}.

\subsection{The Sample of Brown Dwarfs Transiting Main Sequence Stars}

\begin{figure}[t]
\centerline{\includegraphics[angle=0,width=\linewidth]{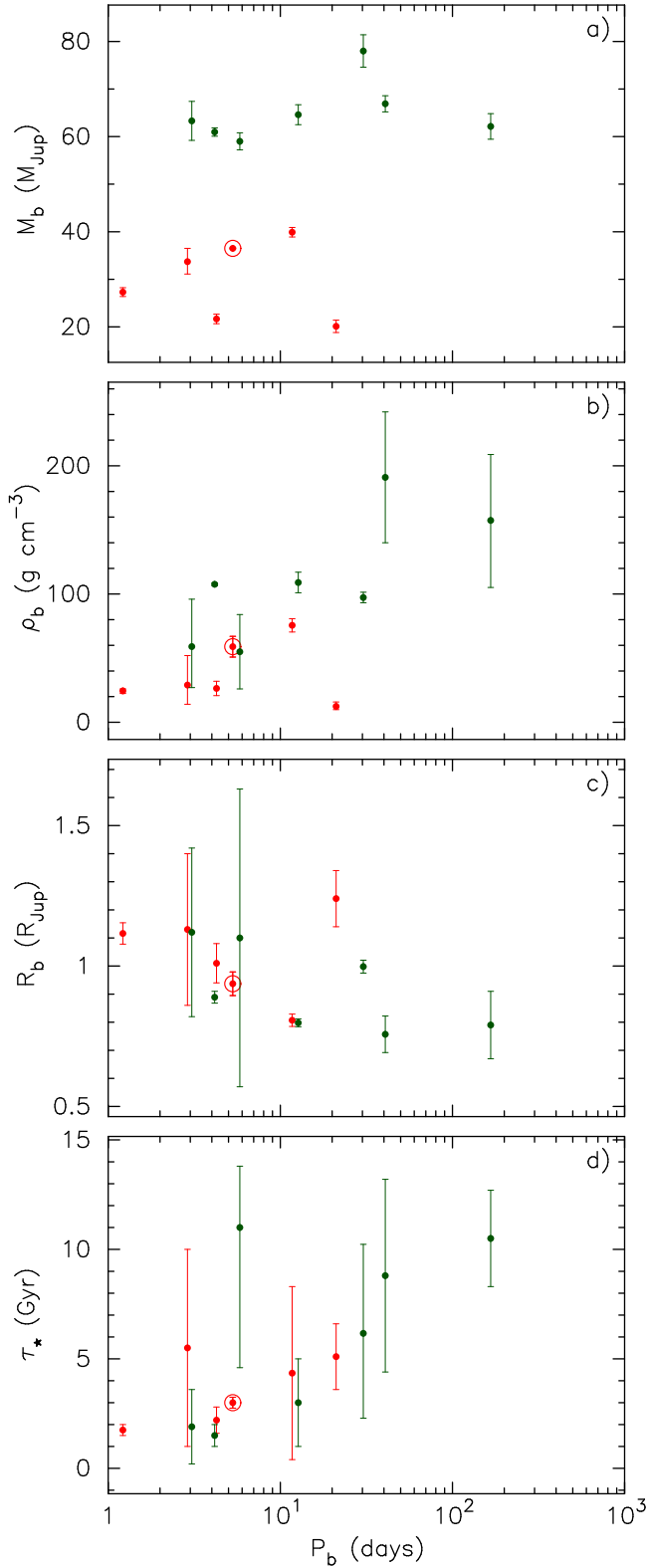}}
\caption{The period-mass, period-density, period-radius and period-stellar age diagram for all brown dwarfs transiting MS stars. Samples and point symbols as for Figure~\ref{figure-10}.
\label{figure-11}}
\end{figure}

The orbital and physical parameters of the sample of known eclipsing brown dwarfs, as well as the atmospheric and physical parameters of their host stars have been recently presented in \cite{2016cole.book..143C}. EPIC\,219388192\,b is the thirteenth brown dwarf found to transit a main sequence star. With an orbital period of 5.3-days and a mass of $M_\mathrm{b}$=$36.500\pm0.090$~$\Mj$, EPIC\,219388192\,b joins the sub-group of six short-period ($P_\mathrm{orb}$\,$<$\,100~days) transiting brown dwarfs with masses below $\sim$45~$\Mj$ (Figure~\ref{figure-11}a). These objects are thought to have formed in the proto-planetary disc through gravitational instability \citep{2014MNRAS.439.2781M}. The other sub-group of brown dwarfs, with masses above $\sim$45~$\Mj$, are believed to have formed via molecular cloud fragmentation. This group consists of seven transiting brown dwarfs, among which six have orbital periods shorter than 100 days.

The group of short-period brown dwarfs less massive than $\sim$45~$\Mj$ is also quite well distinguishable on the period-density diagram (Figure~\ref{figure-11}b), as most of them have densities below 50~$\gpcmcmcm$. With a density of 75.6$\pm$5.6 $\gpcmcmcm$, KOI-205\,b \citep[aka Kepler-492,][]{2013A&A...551L...9D} is the only object above this threshold. Two brown dwarfs more massive than $\sim$45~$\Mj$ (CoRoT-15\,b and CoRoT-33\,b) have densities below 60~$\gpcmcmcm$. These are relatively young objects (Figure~\ref{figure-11}d) still at the beginning of their gravitational contraction. As shown in panel c) of Figure~\ref{figure-11}, there are only two brown dwarfs more massive than $\sim$45~$\Mj$ with radii above 1~$\Rj$, although with large uncertainties. Most of the brown dwarfs with masses smaller than $\sim$45~$\Mj$ have radii below or very close to 1~$\Rj$, and only KOI-205\,b, with a radius of $0.807^{+0.022}_{-0.022}$~$\Rj$, substantially differs from the rest of this group.

EPIC\,219388192\,b is an inhabitant of so called ``brown dwarf desert'' that refers to the paucity of BD companions relative to giant exoplanets within 3~au around MS stars \citep{2000PASP..112..137M,2000A&A...355..581H}. Recently the ``brown dwarf desert'' was however limited only to sub-stellar objects with the $m \sin\,i$ between 35 and 55~$\Mj$ and periods below 100 days \citep{2014MNRAS.439.2781M}. This casts doubts on the proposed distinction between brown dwarfs and planets, which is thought to be connected with different formation mechanisms. \cite{2015ApJ...810L..25H} proposed that objects in the mass range 0.3 - 62 $\Mj$ follow the same relationship on the observed mass-density plot, so they should be considered to belong to one and the same class of celestial objects. Based on planet population synthesis, \cite{2009A&A...501.1139M} showed that the core-accretion mechanism proposed for giant planets formation may produce planets, not only more massive than 13~$\Mj$, i.e. above the deuterium burning limit \citep{2001RvMP...73..719B}, but also in the 20--40~$\Mj$-range. Based on population synthesis calculations of the tidal downsizing hypothesis, \cite{2015MNRAS.452.1654N} have recently suggested that gravitational instability -- proposed as additional formation mechanism to the most natural one for brown dwarfs (via molecular cloud fragmentation) -- can also lead to the formation of giant planets.

The ages of most the stars known to host a transiting brown dwarfs -- with the exception of a few young objects -- are very poorly constrained (Figure~{\ref{figure-11}}d). Any comparisons with theoretical evolutionary models, such as COND03, are therefore limited. More detection of brown dwarfs transiting stars in clusters with robust age determinations are therefore highly desirable to test sub-stellar evolutionary models. Such detection will become possible in some of the upcoming K2 campaigns. The TESS and PLATO space missions monitor large areas of the sky. Including as many open cluster regions as possible in their programmed observations should be a high priority.\\

We are very grateful to the NOT, McDonald, and Subaru staff members for their unique support during the observations. This work was supported by the Spanish Ministry of Economy and Competitiveness (MINECO) through grant ESP2014-57495-C2-1-R. Antonino F. Lanza acknowledge support from the {\it Progetti Premiali} scheme (Premiale WOW) of the Italian national Ministry of Education, University, and Research. This work was supported by the Astrobiology Center Project of National Institutes of Natural Sciences (NINS) (Grant Numbers AB281012 and JY280092). This work was also supported by JSPS KAKENHI (Grant Numbers JP25247026 and JP16K17660). Szilard Csizmadia thanks the Hungarian OTKA Grant K113117. Hans Deeg and David Nespral acknowledge support by grant ESP2015-65712-C5-4-R of the Spanish Secretary of State for R\& D\&i (MINECO). Ignasi Ribas acknowledges support by the Spanish Ministry of Economy and Competitiveness (MINECO) through grant ESP2014-57495-C2-2-R. This research was supported by the Ministerio de Economia y Competitividad under project FIS2012-31079. The research leading to these results has received funding from the European Union Seventh Framework Programme (FP7/2013-2016) under grant agreement No. 312430 (OPTICON) and from the NASA K2 Guest Observer Cycle 1 program under grant NNX15AV58G to The University of Texas at Austin. Based on observations obtained with the Nordic Optical Telescope (NOT), operated on the island of La Palma jointly by Denmark, Finland, Iceland, Norway, and Sweden, in the Spanish Observatorio del Roque de los Muchachos (ORM) of the Instituto de Astrof\'isica de Canarias (IAC). This paper includes data taken at McDonald Observatory of the University of Texas at Austin. This paper includes data collected by the Kepler mission. Funding for the Kepler mission is provided by the NASA Science Mission directorate.


\end{document}